# Spin Glass and Semiconducting Behavior in 1D BaFe$_{2-\delta}$Se$_3$ Crystals

Bayrammurad Saparov[1], Stuart Calder[2], Balazs Sipos[1], Huibo Cao[2], Songxue Chi[2], David J. Singh[1], Andrew D. Christianson[2], Mark D. Lumsden[2], Athena S. Sefat[1]

[1] Materials Science and Technology Division, Oak Ridge National Laboratory, Oak Ridge, Tennessee 37831-6114, USA
[2] Neutron Scattering Science Division, Oak Ridge National Laboratory, Oak Ridge, TN 37831-6393 USA

We investigate the physical properties and electronic structure of BaFe$_{2-\delta}$Se$_3$ crystals, which were grown out of tellurium flux. The crystal structure of the compound, an iron-deficient derivative of the ThCr$_2$Si$_2$-type, is built upon edge-shared FeSe$_4$ tetrahedra fused into double chains. The semiconducting BaFe$_{2-\delta}$Se$_3$ with $\delta \approx 0.2$ ($\rho_{295K}$ = 0.18 $\Omega \cdot$cm and $E_g$ = 0.30 eV) does not order magnetically, however there is evidence for short-range magnetic correlations of spin glass type ($T_f \approx 50$ K) in magnetization, heat capacity and neutron diffraction results. A one-third substitution of selenium with sulfur leads to a slightly higher electrical conductivity ($\rho_{295K}$ = 0.11 $\Omega \cdot$cm and $E_g$ = 0.22 eV) and a lower spin glass freezing temperature ($T_f \approx 15$ K), corroborating with higher electrical conductivity reported for BaFe$_2$S$_3$. According to the electronic structure calculations, BaFe$_2$Se$_3$ can be considered as a one-dimensional ladder structure with a weak interchain coupling.

## I. INTRODUCTION

Following the discovery of superconductivity in fluorine-doped LaFeAsO[1], the iron-based pnictides and chalcogenides have gained a remarkable interest in the condensed matter physics community. The main exploration has been of the arsenides, with efforts focused on two paths: chemical substitutions in the parent compounds and creation of multi-layered compounds featuring the anti-PbO-type FeAs layers. Such work has led to many doped superconductors and also complex unit cells such as Sr$_4$V$_2$O$_6$Fe$_2$As$_2$,[2] which are harder to prepare stoichiometrically, and consequently, can display a variety of physical properties[3]. Compared to the arsenides, the selenide families, especially those with ThCr$_2$Si$_2$ structure-type, are relatively less studied. The insertion of alkali metals in between the FeSe layers in $A_x$Fe$_y$Se$_2$ results in surprising T$_C$ values of $\approx 30$ K[4]. In such intercalation compounds, the alkali metal and iron contents are crucial for superconductivity, and the roles of the disorder or phase separation are also matters of debate[5,6]. For example, high Fe content (1.8 $\leq y$) is important for giving a superconducting state[5,6], intermediate Fe concentrations (1.6 $\leq y \leq$ 1.8) give insulator-metal transitions[5], while lower iron concentrations such as $y$ = 1.6 are insulators[7] with $\sqrt{5} \times \sqrt{5}$ vacancy order[8]. As a part of ongoing investigations on causes of 2D layered structures giving superconductivity and role of iron deficiency in the layers of selenide systems, here is a study of iron-deficient BaFe$_{2-\delta}$Se$_3$ and its partially sulfur-substituted analog.

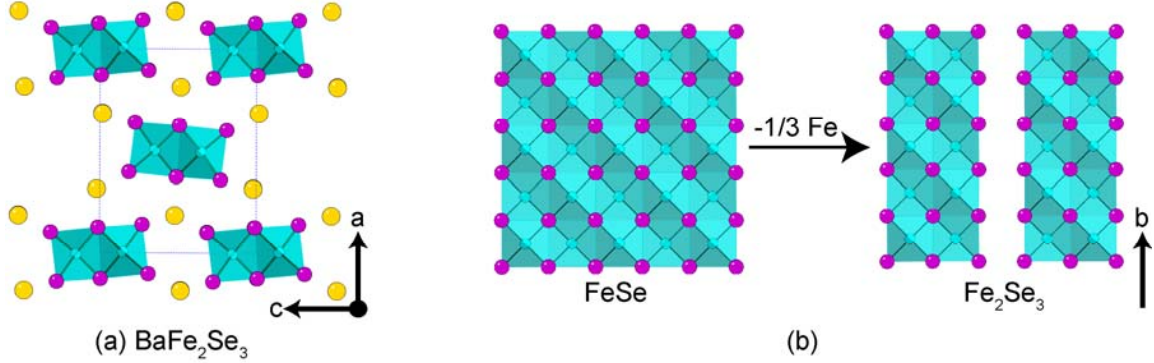

**Figure 1.** (a) The crystal structure of $BaFe_2Se_3$: Ba, Fe and Se atoms are drawn as golden, blue and magenta spheres, respectively. (b) A schematic representation of the relationship between the anionic substructures of the superconducting $A_xFe_ySe_2$ ($ThCr_2Si_2$-type) and $BaFe_2Se_3$: the removal of every third iron atom from [FeSe] layers yields $[Fe_2Se_3]$ double chains.

The $BaFe_2Se_3$ phase is one of only three reported compounds in Ba-Fe-Se ternary system[9]. It crystallizes in the orthorhombic space group *Pnma*[10]. The structure can be described as double chains of $[Fe_2Se_3]^{2-}$ formed by edge-shared $FeSe_4$ tetrahedra extending along the *b*-axis, and $Ba^{2+}$ cations acting as spacers (Figure 1a). The anionic substructure composed of 1D $[Fe_2Se_3]$ chains in $BaFe_2Se_3$ is obtained by removing every third iron atom from the 2D [FeSe] layers in $A_xFe_ySe_2$ ($ThCr_2Si_2$-type) superconductors (Figure 1b). The closely related isoelectronic $BaFe_2S_3$ also features the same structural fragments, *i.e.* tetrahedral double chains separated by cations, however, the double chains $[Fe_2S_3]^{2-}$ are not tilted with respect to each other, resulting in *Cmcm* space group[10]. Investigations on $BaFe_2S_3$ revealed its semiconducting and spin glass behavior below $T_f \approx 25$ K[11,12], whereas recent reports on $BaFe_2Se_3$[13,14] suggest antiferromagnetic order below $T_N = 240$ K[13] (and superconductivity at $T_C = 11$ K) or $T_N = 256$ K[14]. A high temperature superconducting phase is only possible through suppression of an antiferromagnetic order and in order to create such an effect, here we study Fe-deficient $BaFe_2Se_3$. With this, we also test if a superconducting state is possible in a 1D structure. This is a report of thermodynamic and transport property measurements on flux-grown $BaFe_{2-\delta}Se_3$ with $\delta = 0.21(2)$ and crystals partially substituted with sulfur. Instead of superconductivity, we observe spin glass behavior, which is not a great surprise; we explain the experimental results in conjunction with the electronic structure calculations.

## II. EXPERIMENTAL

**Synthesis.** The elemental reactants, dendritic Ba (3N), Fe granules (>3N), Se shot (5N), Te shot (2-5 mm, 6N), and S pieces (> 3N) were purchased from Alfa Aesar. The starting materials were loaded into alumina crucibles in Ba:Fe:S(Se):Te = 1:2:3:15 and Ba:Fe:S:Se:Te = 1:2:1.5:1.5:15 ratios. The alumina crucibles were then placed into silica tubes, which were then sealed under vacuum. The reaction mixtures were heated to 800°C at a rate of 100°C/hour, homogenized at this temperature for 12 hours, cooled to 550°C over 95 hours, then centrifuged to remove tellurium flux. The products contained large needles (up to 9 mm long) of $BaFe_{2-\delta}Se_3$ phase, or smaller thin needles (up to 2 mm long) of $BaFe_{2-\delta}Se_2S$. The reactions aimed at $BaFe_{2-\delta}S_3$ product proved unsuccessful. A faster cooling rate of 10°C/hour afforded slightly smaller crystals of $BaFe_{2-\delta}Se_3$; however, there was no influence of the cooling rate on features in property results.

**Powder X-ray Diffraction and Energy-dispersive X-ray Spectroscopy.** Room temperature X-ray diffraction patterns were collected on a X'Pert PRO MPD X-ray Powder Diffractometer using the Ni-filtered Cu-K$\alpha$ radiation. A typical run aimed at quantitative analysis was in 5-65° (2$\theta$) range with a step size of 1/60° and ~20 seconds/step counting time. Such diffraction patterns were used for phase identification only, which was carried out employing X'Pert HighScore Plus software. The air and/or moisture stability of the compounds were confirmed with scans of the powder X-ray diffraction patterns of the samples left under air for over three weeks. Longer runs aimed at refinements of the crystal structures and chemical compositions were carried out in 10-60° (2$\theta$) range with a step size of 1/60° and a counting time of 330 seconds/step in a continuous scan type. The unit cell parameters and chemical compositions were refined using GSAS[15,16]. Scanning electron microscope and Energy-dispersive X-ray Spectroscopy (EDS) measurements on the single crystals were carried out using a Hitachi-TM3000 microscope with a Bruker Quantax 70 EDS system.

**Single-Crystal X-ray Diffraction.** A single crystal was selected under a microscope and cut to suitable dimensions (0.05×0.05×0.04 mm$^3$). Data were collected on a Bruker SMART APEX CCD-based diffractometer using fine-focus Mo K$\alpha$ radiation ($\lambda$ = 0.71073 Å). A stream of liquid nitrogen was used to cool the crystal to 130(2) K and 230(2) K. The structures were solved by direct methods and refined by full matrix least-squares methods on $F^2$ using the SHELXTL software package[17]. Semi-empirical absorption correction was applied using SADABS within the SHELXTL package. Site occupation factors (SOF) were checked by freeing individual occupancy factors of atom sites. Such procedure revealed that the Fe site is under-occupied (≈ 90%), whereas other sites did not yield deviations above 3$\sigma$. Identical SOF values were determined for another crystal from a different reaction, confirming the reproducibility of this value. All sites were refined anisotropically.

**Physical Property Measurements.** Four-probe electrical resistivity measurements were done on a Quantum Design Physical Property Measurement System (QD PPMS). Selected single crystals were placed on top of a sapphire plate and attached to the sample holder using platinum wires and Dupont 4929N silver paste. Typical measurements were in the 2K to 400K interval both heating up and cooling down under zero and 100 kOe fields. Specific heat $Cp(T)$ measurements were also carried out using the QD PPMS employing the relaxation method in the temperature range of 2K to 60K. The $C/T$ vs $T^2$ plot was fitted at low temperature region to obtain the lattice ($\beta$) and electronic ($\gamma$) contributions. The value of the Debye temperature ($\theta_D$) was calculated using the relationship $\beta = 12\pi^4 R/5\theta^3_D$, with $R$ = 8.314 J/(K·mol).

DC magnetization measurements as a function of temperature and magnetic field were performed using a Superconducting Quantum Interference Device magnetometer. The temperature dependent measurements were carried out in the range of 1.8K to 400K under the applied field of 5 kOe. The field dependent measurements were carried out at 1.8 K and 295 K up to 65 kOe. Such experiments were repeated along the three crystallographic directions of the orthorhombic crystals. To check for possible superconductivity, the crystals were measured in the temperature range of 1.8K to 30K under the applied field of 20 Oe. The polycrystalline samples for magnetization measurements were prepared by grinding the crystals. Time-dependent magnetization behavior of the samples were measured by first cooling them to the target temperatures (40 K and 60 K), turning on the field ($H$ = 10 kOe) for $t_w$ = 60 s, 300 s and 600 s, then turning the field off, and finally measuring the magnetization time dependence. AC susceptibility measurements were carried out using QD PPMS on a polycrystalline sample. In order to obtain a reasonable signal, 7-8 crystals were aligned approximately along *c*-axis. The measurements were carried out in 40-60 K range with 100 Hz and 2000 Hz frequencies and excitation magnetic field set at 10 Oe. No external field was applied.

**Neutron Diffraction.** Single crystal neutron diffraction measurements were performed on the Wide-Angle Neutron Diffractometer (WAND) at the High Flux Isotope Reactor (HFIR) at the Oak Ridge National Laboratory to map out the large accessible range of reciprocal space. WAND uses vertically focused Ge (1 1 3) monochromator that provides a wavelength of 1.48 Å. The crystal was oriented in the (h h l) plane and rotated with respect to the fixed position sensitive detector that covers a scattering angle of 125°. The experiments were carried out at 10 K, 100 K and 300 K. Additionally, single crystal neutron diffraction data was also collected at 4 K and 290 K on the HB-3A four-circle single crystal diffractormeter at HFIR. HB-3A uses a double focusing Si(2 2 0) monochromator with fixed wavelength of 1.536 Å[18]. The data were refined using Fullprof[19] to explore any nuclear and magnetic structure changes between 4 K and 290 K.

**Electronic Structure Calculations.** Density functional calculations were performed using the general potential linearized augmented planewave (LAPW) method[20] as implemented in the WIEN2k computer code[21]. These calculations were performed using the experimental crystal structure of Hong and Steinfink[10] consisting of one dimensional $Fe_2Se_3$ ladders. We used the exchange correlation functional of Perdew, Burke and Ernzerhof[22] with well converged basis sets and Brillouin zone samplings. The LAPW sphere radii were 2.5 bohr, 2.1 bohr and 2.35 bohr for Ba, Fe and Se, respectively. Relativity was included at the scalar relativistic level for the valence states.

## III. RESULTS AND DISCUSSION

**Synthesis and Structure.** The use of metal fluxes and binary self-fluxes for growing ternary iron selenide crystals has been hindered for a few reasons. For example, Sn, a popular flux, forms stable binary SnSe. The use of binary reactive flux FeSe has been limited due to its high melting point. A survey of literature revealed that tellurium has been successfully used for synthesis of single crystals of transition metal chalcogenides such as $MCh_2$ (where $M$ = Ni, Fe, Ru etc., $Ch$ = S, Se, Te)[23]. In fact, the Fe-Te binary phase diagram[24] suggests that tellurium is a good solvent for iron metal. The low melting point of tellurium ($T_{mp} \approx 450$ °C) affords the synthesis at low temperatures, although because of its low boiling point ($T_{bp} \approx 990$ °C), the highest accessible temperature for the reactions using Te as a flux is limited. We successfully used tellurium as flux to grow large single crystals of Fe-deficient $BaFe_2Se_3$ (see below). Our attempts of slow cooling molten reactions (*i.e.* a "modified" Bridgman technique)[13,14] yielded products containing impurity phases such as FeSe, $FeSe_2$ and Fe as already claimed in previous reports[13,14], in addition to $Fe_7Se_8$ and traces of elemental Se. Our inability to synthesize pure single crystals of required dimensions and quality using the stoichiometric melts prevented further studies of these samples.

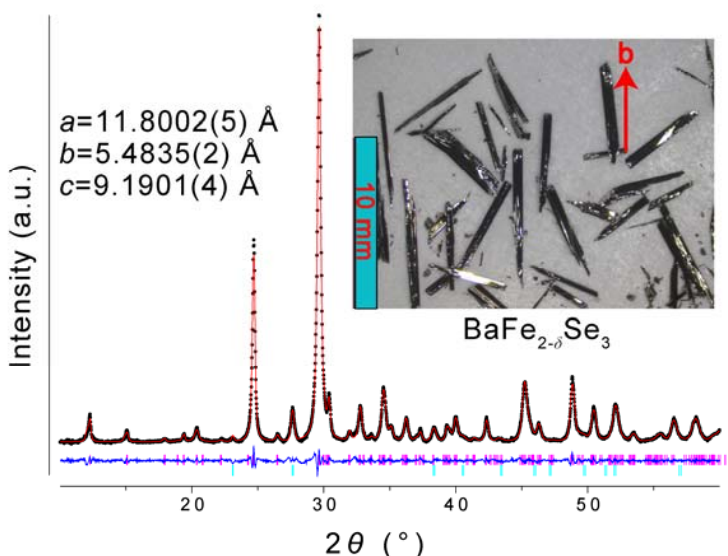

**Figure 2.** Powder X-ray diffraction pattern for powders of Fe-deficient $BaFe_2Se_3$ crystals (black dots). The Te diffraction peaks, source of which is crystal surface flux, are shown in light blue markers; $BaFe_2Se_3$ phase peaks are shown in pink tickmarks. The Rietveld fit and difference plot are shown as red and dark blue lines, respectively. The inset shows a typical crystal yield from a reaction with a total mass 2.5 g including the flux.

For Fe-deficient $BaFe_2Se_3$, a powder X-ray diffraction pattern, along with an image of a typical batch of crystals, are provided in Figure 2. The unit cell parameters were refined to $a = 11.8002(5)$ Å, $b = 5.4835(2)$ Å, and $c = 9.1901(4)$ Å with a reliability factor of $Rp = 1.06\ \%$ (weighted $R$-factor $wRp = 1.44\ \%$) and $\chi^2$ (goodness-of-fit) = 3.01. The refinements did not indicate any Te-flux substitution on the selenium sites (Wyckoff position 4$c$). The absence of Te inclusion was also supported by the EDS measurements on several crystals (more than 20 measurements in total), although several crystals had patches of residual flux on the surfaces (see Figure 2 inset). The diffraction pattern of sulfur-substituted $BaFe_2Se_3$ powder was refined to $a = 11.5962(6)$ Å, $b = 5.3679(3)$ Å, and $c = 9.0468(4)$ Å, with $Rp = 1.52\ \%$ ($wRp = 2.21\ \%$) and $\chi^2 = 3.83$. The unit cell parameters reported for $BaFe_2S_3$ by Steinfink *et al.* are $a = 8.7835(9)$ Å, $b = 11.219(1)$ Å, and $c = 5.2860(5)$ Å[10]. As expected, unit cell volume for the sulfur-substituted $BaFe_2Se_3$ ($V = 563.14(5)$ Å$^3$) falls in between that of the end members, $BaFe_{2-\delta}Se_3$ ($V = 594.66(4)$ Å$^3$) and $BaFe_2S_3$ ($V = 520.89(9)$ Å$^3$). This suggests substitution of one third of selenium with sulfur, if direct proportionality is assumed in the series and Vegard's law is applied. In fact, the EDS measurements on six different single crystals from the $BaFe_2Se_{3-x}S_x$ batch yielded $Ba_{1.01(7)}Fe_{2.00(7)}Se_{2.08(9)}S_{1.0(1)}$ formula.

The sulfur-substituted analog was synthesized with the aim of uncovering the physical property trends in the $BaFe_2Se_3$-$BaFe_2S_3$ phase diagram. The sulfur substitution reduced the crystal size and quality, and we did not find the study of solid solution worthwhile. In fact, the end member $BaFe_{2-\delta}S_3$ could not be synthesized using Te-flux and the highest sulfur solubility achieved corresponds to a formula of $BaFe_{2-\delta}Se_2S$, with much smaller crystals compared to $BaFe_{2-\delta}Se_3$. Among the reasons for this, parameters such as solubility of respective elements in flux and presence of different competing phases in two systems may be limiting factors.

**Table 1.** The unit cell dimensions of $BaFe_{2-\delta}Se_3$ obtained using powder (PXRD) and single crystal X-ray diffraction (SXRD), and neutron powder diffraction (NPD). The reported values in literature and this work are compared.

| $a$, Å | $b$, Å | $c$, Å | $V$, Å$^3$ | T, K | Method | Reference |
|---|---|---|---|---|---|---|
| 11.8903(1) | 5.40798(5) | 9.1375(1) | 587.56(1) | 2 | NPD | (13) |
| 11.8834 | 5.4141 | 9.1409 | 588.106 | 5 | NPD | (14) |
| 11.878(3) | 5.447(2) | 9.160(2) | 592.6(3) | RT | SXRD | (10) |
| 11.757(2) | 5.4445(9) | 9.150(2) | 585.7(2) | 130 | SXRD | This work |
| 11.776(2) | 5.457(1) | 9.158(2) | 588.5(2) | 230 | SXRD | This work |
| 11.8002(5) | 5.4835(2) | 9.1901(4) | 594.66(4) | RT | PXRD | This work |

The reported lattice parameters for $BaFe_2Se_3$ differ significantly between the literature[10,13,14] (Table 1). Such discrepancies in the unit cell parameters reported for different samples could, in fact, be indicative of vacancies in the crystal structure. If stoichiometric $BaFe_2Se_3$ composition is assumed in all of the samples studied in literature, the temperature dependence of the cell dimensions immediately draws attention (Table 1[10,13,14]). As expected, the unit cell volume, and the $b$- and $c$-parameters increase with increasing temperature. However, the $a$-axis undergoes a negative thermal expansion over 3 K and also 290 K temperature intervals (Table 1). The unit cell dimensions reported in this work are significantly lower than the previous reported values. The single crystal X-ray diffraction measurements (see below) on a crystal of $BaFe_{1.79(2)}Se_3$ at 130 K and 230 K rule out a possible negative thermal expansion of the $a$-axis, and therefore, the observed discrepancies in the unit cell parameters are due to the difference in compositions of these samples. The composition of one third substituted selenide-sulfide sample has been only verified using the EDS measurements (see above), which indicates a ratio of Se:S = 2.08(9):1.0(1)

**Table 2.** Atomic coordinates and equivalent isotropic displacement parameters ($U_{eq}^a$) for $BaFe_{1.79(2)}Fe_2Se_3$ at 230(2) K.

| Atom | Wyckoff Position | $x$ | $y$ | $z$ | $U_{eq}$, Å$^2$ | SOF |
|---|---|---|---|---|---|---|
| Ba | 4c | 0.18519(8) | 0.25 | 0.5214(1) | 0.0392(3) | 1 |
| Fe | 8d | 0.4940(1) | 0.0013(2) | 0.3499(1) | 0.0176(4) | 0.89(1) |
| Se1 | 4c | 0.3576(1) | 0.25 | 0.2279(1) | 0.0261(3) | 1 |
| Se2 | 4c | 0.62216(9) | 0.25 | 0.4925(1) | 0.195(3) | 1 |
| Se3 | 4c | 0.3982(1) | 0.25 | 0.8142(1) | 0.0282(3) | 1 |

$^a U_{eq}$ is defined as one third of the trace of the orthogonalized $U_{ij}$ tensor.

In order to determine the exact chemical composition of the Fe-deficient crystals and explain the existing discrepancy in the unit cell parameters, we carried out single crystal X-ray diffraction experiments. The unit cell dimensions at 130 K and 230 K are presented in Table 1; the final positional and equivalent isotropic displacement parameters at 230 K are listed in Table 2. The anisotropic displacement parameters (ADP) are elongated for Ba, Se1 and Se3 sites. The residual peak of 6-7 e$^-$/Å$^3$ is located ≈ 0.78 Å away from Ba position, and is a result of the motion of Ba atoms in the lattice. The final $R$-values are $R_1$ = 4.52%, $wR_2$ = 10.30% (Goodness-of-fit = 1.091) and $R_1$ = 4.36%, $wR_2$ = 9.94% (Goodness-of-fit = 1.085) for 130 K and 230 K data, respectively. Based on these, the Ba atoms and [Fe$_2$Se$_3$] chains appear to be loose in BaFe$_2$- $_\delta$Se$_3$. The iron site is not fully occupied (site occupancy factor SOF = 0.79(2)) and no tellurium mixing on selenium sites occurs resulting in a chemical composition of BaFe$_{1.79(2)}$Se$_3$. Consequently, the iron atoms in BaFe$_{1.79(2)}$Se$_3$ are expected to display mixed valence in order to maintain the charge balance. In agreement with this finding, "BaFe$_2$S$_3$" has already been characterized as mixed valence compound in literature[25]. The ADP values reported by Steinfink et al.[10] also show enlarged Ba, Se1 and Se3, and Ba and S2 sites in BaFe$_2$Se$_3$ and BaFe$_2$S$_3$, respectively. The reported reliability factors are $R1$ = 6.4% for BaFe$_2$S$_3$ and $R1$ = 8.91% for BaFe$_2$Se$_3$, however, no partial occupancy of iron site was reported. Other reports of single crystal X-ray investigations on BaFe$_2$Se$_3$ and BaFe$_2$S$_3$ can be found in literature[12,13], although no details of the obtained results, crystallographic data, ADPs, SOFs or unit cell parameters were provided, making a direct comparison between the data sets impossible. In the further sections, a chemical formula of BaFe$_{1.79(2)}$Se$_3$ is implied for BaFe$_{2-\delta}$Se$_3$. For the sulfur-substituted crystals, a formula of BaFe$_{2-\delta}$Se$_2$S is implied based on the EDS measurements, where under-occupancy of the iron site is similarly expected.

**Physical Properties.** Electrical resistance data as a function of temperature for BaFe$_{2-\delta}$Se$_3$ and BaFe$_{2-\delta}$Se$_2$S are plotted in Figure 3. The resistance increases with decreasing temperature ($\approx$ 6 orders of magnitude change in 300 K range), indicating insulating behavior. The resistance along and perpendicular to the needle propagation directions are identical within experimental error, suggesting no anisotropy of the electrical transport properties along $b$- and $a,c$-axes. The room temperature resistivity $\rho_{295K}$ values for BaFe$_{2-\delta}$Se$_3$ and BaFe$_{2-\delta}$Se$_2$S are 0.18 $\Omega$·cm and 0.11 $\Omega$·cm, respectively. The substitution of selenium with sulfur is expected to make the material more insulating due to higher electronegativity of sulfur, however, our data show the reverse trend. Notwithstanding this fact, the numerical value of the electrical resistivity of BaFe$_2$S$_3$ ($\rho_{295K} \approx$ 0.09 $\Omega$·cm along $b$-axis)[11] at each temperature is lower than that of BaFe$_{2-\delta}$Se$_3$ as well, indicating the more conductive nature. The resistivity data obtained for BaFe$_{2-\delta}$Se$_3$ is in an agreement with the measurements reported for the related K$_x$Fe$_y$Se$_2$ (ThCr$_2$Si$_2$-type) series as well, according to which the iron content of $y \leq$ 1.6 results in insulating behavior of compounds[5,7]. Indeed, the iron content in BaFe$_2$Se$_3$ could be shown as Fe$_{2/1.5}$Se$_{3/1.5}$ = Fe$_{1.33}$Se$_2$, hence, $y$ is less than 1.6 and the compound is insulating.

Arrhenius plot for BaFe$_{2-\delta}$Se$_3$ gives two linear regions for 100-160 K and 250-380 K with the calculated band gaps ($E_g$) of 0.24 eV and 0.30 eV, respectively. These results are in an agreement with the reported temperature dependence of the electrical resistivity of BaFe$_2$S$_3$[11], which also exhibits two linear regions with $E_g$ = 0.16 eV ($\approx$ 250-400 K region) and 0.07 eV ($\approx$ 70-250 K region). Interestingly, Eichhorn et al.[12] report an almost identical band gap for BaFe$_2$S$_3$, although the numerical values for resistivity are much lower (the values based on the extrapolations of the graphs in the two papers differ by 3 orders of magnitude) compared to that reported by Steinfink et al.[11] According to this report[12], BaFe$_2$S$_3$ crystals demonstrate a negative magnetoresistive effect of 10% at 90 kOe. No magnetoresistive behavior was observed for BaFe$_{2-\delta}$Se$_3$ under 100 kOe in 400 K to 100 K temperature range. For BaFe$_{2-\delta}$Se$_2$S, a band gap of $E_g$ = 0.22 eV was obtained from the fit in 250-380 K region. The calculated band gap and electrical resistivity for BaFe$_{2-\delta}$Se$_2$S falls in between that for BaFe$_{2-\delta}$Se$_3$ and BaFe$_2$S$_3$.

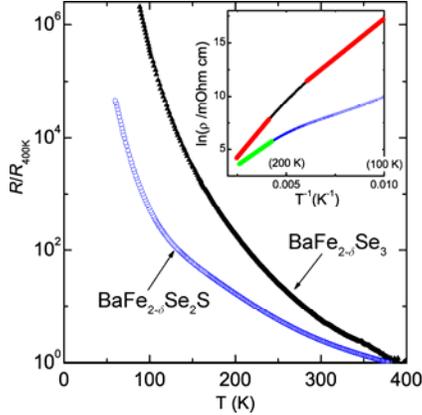

**Figure 3.** Plot of the temperature dependence of the electrical resistivities of BaFe$_{2-\delta}$Se$_3$ and BaFe$_{2-\delta}$Se$_2$S. The inset shows linear red and green fits according to the Arrhenius temperature dependence of the resistivity $\ln \rho = \ln \rho_0 - E_g / 2k_B T$.

The temperature dependence of the magnetic susceptibility (Figure 4a) for BaFe$_{2-\delta}$Se$_3$ is anisotropic. There is a clear peak at ≈ 44 K in zero-field cooled data perpendicular to the chain propagation directions ($\parallel a$ and $\parallel c$). Below this temperature, there is a divergence in the field-cooled (FC) and the zero field-cooled (ZFC) data both along *a*- and *c*-axes, but not the *b*-axis. Beyond 44 K, there are no significant features. Also, there is no Curie-Weiss behavior up to 400 K (Figure 4a, inset). The ZFC magnetization data for polycrystalline BaFe$_{2-\delta}$Se$_3$ and BaFe$_{2-\delta}$Se$_2$S samples are displayed in Figure 4b. The cusp is at a lower temperature of 15 K for BaFe$_{2-\delta}$Se$_2$S compared to BaFe$_{1.79(2)}$Se$_3$. The absence of superconductivity in BaFe$_{2-\delta}$Se$_3$ was confirmed (Figure 4c) by DC magnetization experiment under applied field of 20 Oe. The M(H) plot (Figure 4d) along *b*-axis shows almost linear dependence with small curvature noticeable after 20 kOe at 1.8 K. Along *a*- and *c*-axes, the *M(H)* dependence for BaFe$_{2-\delta}$Se$_3$ shows hysteresis (Figure 4d, inset) with a small remanent moment ($M_r$) of 0.025 emu/g and a coercive field ($H_c$) of 2 kOe. For BaFe$_{2-\delta}$Se$_2$S (not shown), slightly lower values of $M_r$ = 0.020 emu/g and $H_c$ = 1 kOe were determined. The field-dependent magnetization measurements at room temperature (295 K) show linear dependence for all directions with no hysteresis. The divergence in ZFC/FC data usually indicates the presence of a ferromagnetic (FM) signal, but *M(H)* data is linear and does not saturate up to 65 kOe. According to this data, we suspect that the compounds are canted antiferromagnets (AFM) or spin glasses deriving small FM moments.

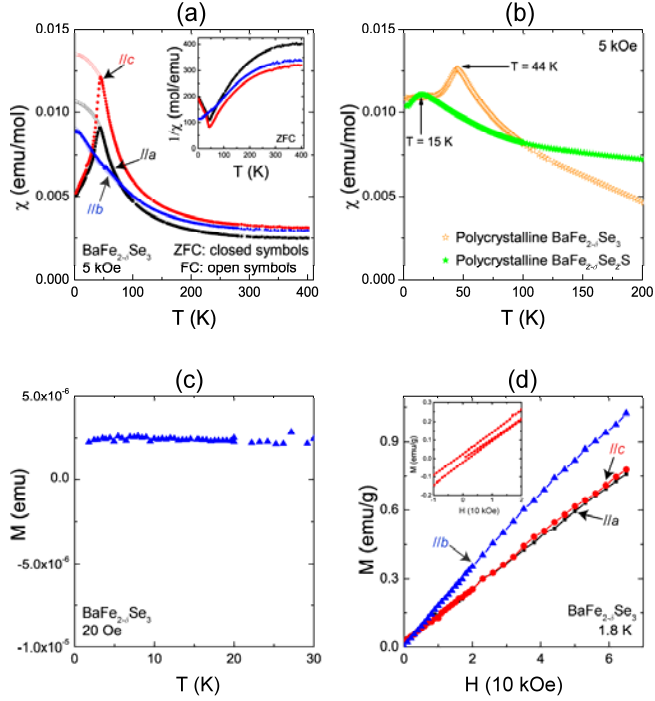

**Figure 4.** (a) Temperature dependence of the magnetization for BaFe$_{2-\delta}$Se$_3$ measured along *a*-, *b*- and *c*-axes under 5 kOe. The inset shows the reverse of the zero-field cooled susceptibility data *vs* T. (b) Temperature dependence of the magnetization (ZFC) for BaFe$_{2-\delta}$Se$_3$ and BaFe$_{2-\delta}$Se$_2$S measured on polycrystalline samples. (c) ZFC magnetic susceptibility of BaFe$_{2-\delta}$Se$_3$ measured under the applied field of 20 Oe. (d) The magnetization versus field at 1.8 K for BaFe$_{2-\delta}$Se$_3$ along all crystallographic directions. The inset shows a weak hysteresis in the *M(H)* plot along *c*-axis.

Our magnetization data do not show superconductivity, like that reported by Conder *et al.*[13] at 11 K, see Figure 4c. We also do not have long range order as reported below 240 K[13] or 256 K[14]. Our data corroborate with the results of the related BaFe$_2$S$_3$[12] with a spin glass behavior. For BaFe$_2$S$_3$, no long range order at high temperatures (200-300K) was observed, instead, a cusp feature was found at 25 K[12], similar to our BaFe$_{2-\delta}$Se$_3$ and BaFe$_{2-\delta}$Se$_2$S data. The sulfur analog, BaFe$_2$S$_3$, also demonstrated a small hysteresis with $M_r$ = 0.0623 emu/g and $H_c$ = 1.8 kOe[12]. Furthermore, the magnetization for BaFe$_2$S$_3$ does not saturate in the measured range (up to 60 kOe)[12].

To confirm spin glass behavior in BaFe$_{1.79(2)}$Se$_3$, we carried out AC susceptibility measurement at two frequencies (Figure 5a). The results showed there is a shift in the glass freezing temperature ($T_f$) with frequency. Such frequency dependence is an indication of slow spin dynamics associated with the spin-glass freezing temperature. It should be noted that, similar to BaFe$_2$S$_3$[12], the change in $T_f$ is minimal as the frequency is changed. The spin-glass behavior of BaFe$_{2-\delta}$Se$_3$ has been further verified through the time-dependent magnetization decay measurements (Figure 5b). The magnetic relaxation in spin-glasses may be fitted to the *stretched exponential equation*, $M = M_2 + (M_1 - M_2)\exp\left[-\left(\dfrac{t}{\tau}\right)^\beta\right]$, where $M_1$ and $M_2$ are the initial and final magnetization, respectively; $\tau$ is the relaxation time constant and $\beta$ can take values from 0 (no relaxation) to 1 (exponential Debye relaxation). The terms $M_2$ and ($M_1$-$M_2$) relate to an intrinsic ferromagnetic component and a glassy component, respectively. The term $t_w$ is the time BaFe$_{2-\delta}$Se$_3$ is kept under the field before the measurement. The fit of the 40 K data to the equation above gives $\beta = 0.67$ for all three $t_w$, while the time constant increases from $\tau = 671$ s for $t_w = 60$ s to $\tau = 710$ s for $t_w = 600$ s. Such changes in the $M(t)$ for different values of $t_w$ is an indication of aging effects in a metastable spin glass compound[26]. The reported values of $\beta$ and $\tau$ are 0.54 and 1200, respectively, for BaFe$_2$S$_3$[12]. At a temperature above $T_f$ (60 K) no exponential decay of magnetization was observed (Figure 5).

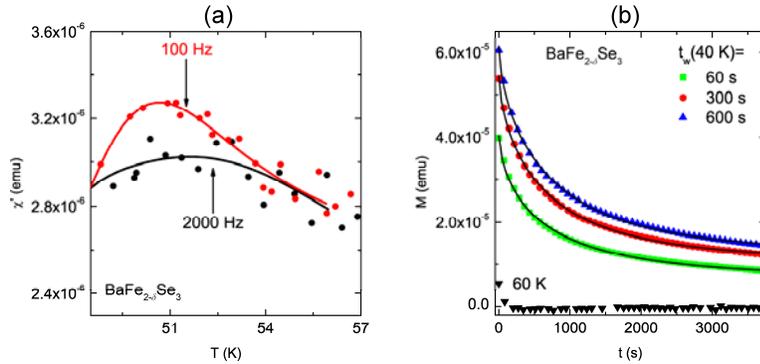

**Figure 5.** (a) Temperature dependence of the real AC magnetic susceptibility ($\chi'$) component. Temperature sweeps were conducted at 100 Hz and 2000 Hz frequencies. (b) Remanent DC magnetization relaxation for BaFe$_{2-\delta}$Se$_3$ at $T = 40$ K ($t_w = 60$ s, 300 s and 600 s) and 60 K ($t_w = 300$ s) measured along $c$-axis. The 40 K data are fitted to the stretched exponential equation (see text).

The specific heat data for BaFe$_{2-\delta}$Se$_3$ is shown in Figure 6. A careful scan of the 39-49 K region revealed no phase transition around 44 K, which is consistent with the absence of long-range order. From the fit of C/T versus T$^2$ (Figure 6, inset) below 8 K, the electronic and lattice contributions were estimated to be $\gamma \approx 0$ for the Sommerfield coefficient and $\beta = 0.00044$ J mol$^{-1}$ K$^{-2}$ atom$^{-1}$, respectively. The latter was used to estimate the Debye temperature of $\theta_D \approx 149$ K. While the obtained value for $\gamma$ is expected for an insulator, the calculated Debye temperature is much lower than the reported value of $\theta_D = 435$ K for BaFe$_2$S$_3$[27].

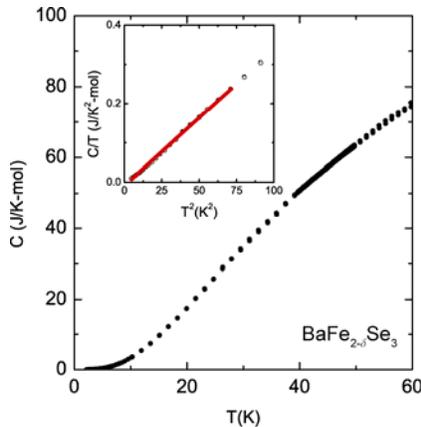

**Figure 6.** The temperature dependence of specific heat for BaFe$_{1.79(2)}$Se$_3$. Plot of C/T vs T$^2$ below 10 K is shown in the inset.

**Results of Neutron Scattering Experiments.** In addition to heat capacity evidence, the absence of long range antiferromagnetic order is further supported by the neutron diffraction experiments seen in Figure 7. Bragg peaks can be clearly seen at the expected integer positions at all temperatures investigated. No additional scattering is observed at half-integer values in cooling from 300 K to 10 K, contradicting to the reported **k** = (1/2 1/2 1/2) magnetic structure[13,14]. Additionally, there is no increase in scattering on nuclear Bragg peaks and no additional peaks in the (H H L) plane are observed with decreasing temperature that would also indicate magnetic order. Therefore, from these results, there is no detectable evidence of long range magnetic order below 300 K in our $BaFe_{2-\delta}Se_3$ crystals.

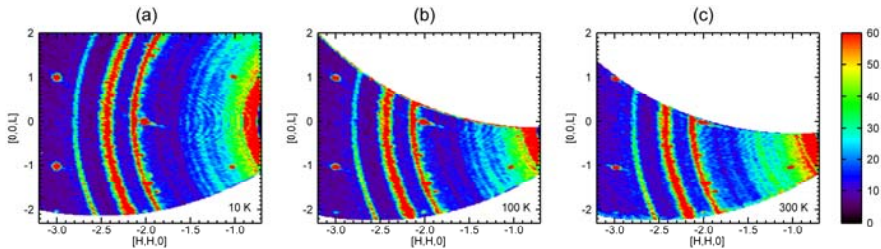

**Figure 7.** The observed neutron diffraction pattern in the (H,H,L) scattering plane (a) 10 K, (b) 100 K and (c) 300 K obtained on WAND, with the scale representing the intensity of scattering. The continuous high intensity scattering corresponds to the expected powder rings from the Al sample environment.

The data obtained on the HB-3A four-circle single crystal diffractometer is presented in Figure 8. There is clearly no change in intensity between 290 K and 4 K expected for a **k** = (1/2 1/2 1/2) magnetic structure[13,14] (Figure 8a). In fact, the only noticeable change in the intensity is in peaks such as (8 0 1), (4 0 1) etc. (Figure 8b). The (8 0 1) peak was chosen for further study, and it seems to show a continuous increase upon cooling (Figure 8c). This is indicative of a continuous structure variation. The refinement of the two data collections at 4 K and 290 K imply small rotation of the $[Fe_2Se_3]$ chains along the *b*-axis when cooling, which might be related to the cause of the observed spin-glass behavior of the compound. To study this continuous change in the local structure, more reflections are needed and such work is currently underway. Based on the data obtained here, no measurable ordered moment is found.

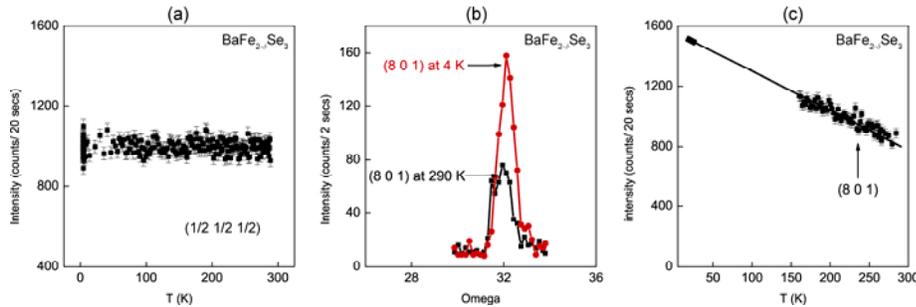

**Figure 8.** The neutron diffraction data obtained on the four-circle single crystal neutron diffractometer. (a) A scan from 290 K down to 4 K showing no diffraction peaks with a **k** = (1/2 1/2 1/2) wave vector. (b) Comparison of the intensities of (8 0 1) peak at 4 K and 290 K. (c) Temperature dependence of the intensity of (8 0 1) peak upon cooling.

**Electronic Structure Calculations.** The electronic structure calculations were performed in order to gain an insight into the observed electrical and magnetic properties of this compound, although modeling of a sample without a long range order is impossible. Comparisons of the results of such calculations for different models could give an indication of strength of intra- and inter-chain interactions. With these ideas in mind, the electronic structure calculations were first performed for a non-spin-polarized case, which we find to be metallic but strongly unstable against moment formation on the Fe sites. The electronic structure for this non-spin-polarized metal is very distinct from that of the Fe superconductors[28,29], and in particular, it does not show a dip in the density of states around the Fermi energy corresponding to the small Fermi surfaces thought to be of importance to the superconductivity of those compounds. This suggests that $BaFe_2Se_3$ is not related to the Fe superconductors in an electronic sense.

**Table 3.** Calculated spin moments (m) and energies on a per Fe basis relative to the non-spin-polarized case (E) for different magnetic orderings of $BaFe_2Se_3$ (see text).

| Model | E (meV/Fe) | m ($\mu_B$) |
|---|---|---|
| PM | 0 | 0.00 |
| FM | -364 | 2.87 |
| AF-XY | -370 | 2.89 |
| AF-P | -532 | 2.84 |

As mentioned, the non-spin-polarized state (PM) is strongly unstable against moment formation. We did calculations for three different magnetic orders, specifically, ferromagnetic (FM), an antiferromagnetic state consisting of ladders with all Fe moments in a ladder ferromagnetically aligned, but with neighboring ladders opposite to each other (AF-XY) and the antiferromagnetic state proposed in recent neutron experiments[14], consisting of four Fe atom plackets along the ladders (AF-P). The energies and Fe spin moments defined by the spin polarization inside an Fe LAPW sphere are given in Table 3. As seen in Table 3, the Fe moments are practically the same to within 0.05 $\mu_B$ for the different orderings and furthermore the energy differences between the various magnetic orderings are all substantially smaller than the energy of the non-spin-polarized state (PM). This shows local moment magnetism. Comparing the energy of the FM and AF-P states (difference of 168 meV/Fe; if the actual ground state differs from AF-P, its energy below the FM state will be even larger), one notes that the magnetic ordering energy scale is very high, which would normally imply a high ordering temperature (but see below) and high spin wave velocities along the ladders in the ordered state.

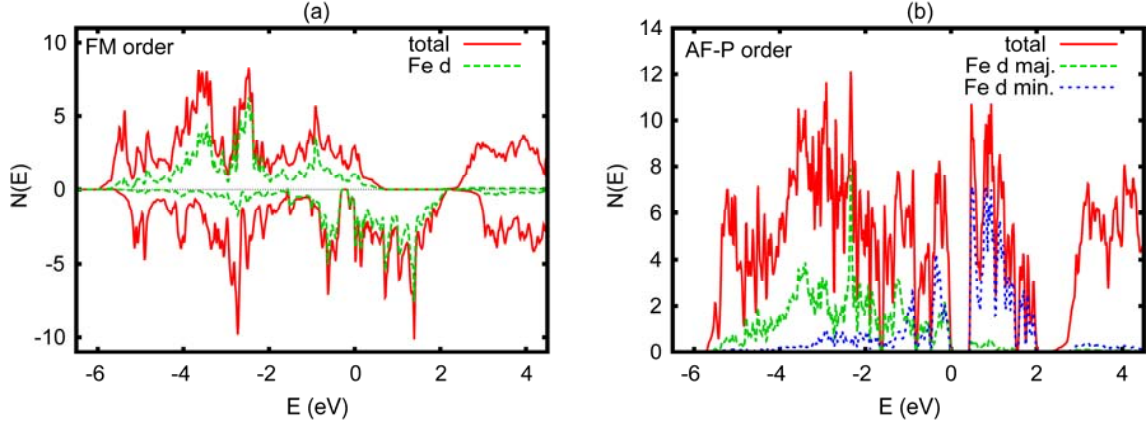

**Figure 9.** (a) Majority (above axis) and minority (below axis) total and Fe $d$ electronic density of states with FM ordering on a per formula unit basis. The Fermi energy is at 0 eV. (b) Total and majority and minority Fe $d$ projections of the density of states for the AF-P ordering. The valence band maximum is at 0 eV.

The FM ordering also is metallic as is the AF-XY. The calculated electronic density of states with FM ordering is shown in Figure 9a along with the Fe $d$ projection for majority and minority spin. There is a strongly spin dependent hybridization between Fe $d$ states and Se $p$ states. Furthermore the majority spin Fe $d$ states are fully occupied. As shown in Table 3, the moments inside the Fe LAPW sphere amount to 2.87 $\mu_B$ per Fe. The total spin magnetization, including small polarizations on Se and in the interstitial as well, amounts to 3.37 $\mu_B$ per Fe. This is less than the full high spin value for $Fe^{2+}$ reflecting the effect of hybridization with Se. Considering the fact that the bands over the whole valence band region down to 6 eV binding energy contain substantial mixtures of Fe and Se character, the anionic ladders in $BaFe_2Se_3$ may be regarded as at least as covalent and perhaps more so than the FeSe sheets comprising the superconductor FeSe.

The AFM-XY differs from the FM ordering in that it has the maximum possible number of antiferromagnetic Fe-Fe pairs between different neighboring ladders, while the FM order has zero. The very small energy difference between these orders – only 6 meV/Fe in spite of the high Fe – moments is indicative of very weak coupling between different ladders. This is particularly striking when compared to the large energy difference involved in the intra-layer ordering – the 168 meV/Fe difference between FM and AF-P. This means that from a magnetic point of view $BaFe_2Se_3$ is properly described as one dimensional ladders with very weak interactions between them. Since such weak interactions between anionic units generally arise from overlap of the tails of wavefunctions associated with those units, we expect these inter-ladder interactions to be very sensitive to structure (overlaps of wavefunction tails are generally exponential in the bond lengths). The implication is that there will be a suppression of long range antiferromagnetic order due to dimensionality and a strong sensitivity of the ordering to the structural and chemical details of the sample in question.

The calculated electronic density of states and projections for the AF-P state are shown in Figure 9b. As may be seen this compound has essentially fully occupied majority spin Fe $d$ states, and a strong spin dependent hybridization, which reduces the moment as in the FM case. However, unlike the FM case, this ordering produces an insulating ground state with a band gap of $E_g = 0.44$ eV. Assuming that the compound is a d-d magnetic band insulator, as would be reasonable for such a covalent non-oxide material; this would be the predicted gap. Presumably this gap will persist even if the ladders are not ordered, as long as they have local short-range antiferromagnetic order along the ladder direction. For a short range order (spin glass) along the ladders, sheets of diffuse magnetic scattering in neutron diffraction are expected to be observed. The study of development of such sheets will be useful in understanding the temperature evolution of the short range order.

## IV. CONCLUSIONS

The compound BaFe$_{2-\delta}$Se$_3$ does not show long range magnetic order in the 1.8 K to 400 K temperature range. A divergence of the zero field-cooled and field cooled data is observed at 44 K along the *a*- and *c*-axes. The field dependent magnetization data perpendicular to the *b*-axis shows hysteresis at low temperatures (1.8 K) with a remanent moment of 0.025 emu/g and a coercive field of 2 kOe. The magnetization experiments, including the time-dependent DC magnetization decay experiments and AC magnetic susceptibility measurements point to the presence of short-range magnetic correlations in the structure and an anisotropic spin glass behavior for BaFe$_{2-\delta}$Se$_3$. The absence of long-range magnetic order is further supported by the no-feature result in specific heat data. The neutron diffraction experiments confirm the property measurements data, as only Bragg peaks and no reflection originating from a magnetic structure were observed. The electronic structure calculations reveal weak interladder coupling (interactions along *a*- and *c*-axes) and one-dimensional nature of the compound. This, along with the vacancies in the iron site, could explain spin glass behavior of BaFe$_{2-\delta}$Se$_3$.

The semiconducting behavior of BaFe$_{2-\delta}$Se$_3$, an iron deficient derivative of $A_x$Fe$_y$Se$_2$ superconductors, is in line with the earlier reports of detrimental influence of iron deficiency to the superconductivity[5-7]. BaFe$_{2-\delta}$Se$_3$ has a band gap of $E_g$ = 0.30 eV and $\rho_{295K}$ = 0.18 Ω·cm. The slope of ln$\rho$ versus T$^{-1}$ changes below approximately 250 K resulting in a band gap of 0.24 eV, and from specific heat, the Sommerfield coefficient $\gamma$ is ≈ 0. No magnetoresistance has been observed for BaFe$_{2-\delta}$Se$_3$ under 100 kOe down to ≈ 100 K. However, the reported negative magnetoresistance of 10% for BaFe$_2$S$_3$ was observed below $T_f$ = 25 K. Because our samples are too resistive to measure below 100 K, the magnetoresistance of BaFe$_{2-\delta}$Se$_3$ could not be confirmed below $T_f$ ≈ 50 K. A sulfur-substituted analog of the phase, BaFe$_{2-\delta}$Se$_2$S, has a more-disordered structure and therefore, the cusp feature in magnetization data at $T_f$ = 15 K is lower than both that of BaFe$_{2-\delta}$Se$_3$ and BaFe$_2$S$_3$. BaFe$_{2-\delta}$Se$_2$S shows smaller electrical resistivity and band gap values of $\rho_{295K}$ = 0.11 Ω·cm and $E_g$ = 0.22 eV, compared to BaFe$_{1.79(2)}$Se$_3$. This somewhat surprising trend in electrical resistivity data in this family is corroborated by the earlier studies of the temperature dependence of the electrical resistivity of BaFe$_2$S$_3$ with the reported values of $E_g$ = 0.16 eV and $\rho_{295K}$ ≈ 0.09 Ω·cm[11].

The earlier reports on BaFe$_2$Se$_3$ suggest long range antiferromagnetic order at $T_N$ = 240 K[13] or $T_N$ = 256 K[14], each associated with a different magnetic structure. Because the reported samples were prepared from stoichiometric melts and presence of impurity phases has been documented, their observed properties could also have been influenced by off-stoichiometry (*i.e.* BaFe$_{2-x}$Se$_{3-y}$), giving slightly different $T_N$ values with different magnetic ground states. For example, presence of 1% Fe impurity only[14] necessitates vacancies in iron sites in BaFe$_2$Se$_3$, whereas FeSe impurity[13] requires vacancies both in iron and selenium sites. Difference in chemical compositions in these samples is further illustrated by the apparent inconsistency of the unit cell parameters. It is somewhat surprising that the reported antiferromagnetic structures[13,14] feature ferromagnetic Fe$_4$ plaquettes, though arranged differently, instead of completely anti-aligned spins in the ladders, considering the short Fe-Fe distances in the [Fe$_2$Se$_3$] double chains.

Based on the obtained data so far, higher concentrations of iron vacancies may lead to a disordered spin glass behavior, while lower concentrations of iron vacancies may lead to a long range antiferromagnetic order. Consequently, one could expect an even higher $T_N$ value for stoichiometric $BaFe_2Se_3$, which in turn could feature completely anti-aligned spins in the chains. Our result on the spin glass behavior of $BaFe_{2-\delta}Se_3$ is another testament to the idea that superconductivity may only be possible in structures with 2D layers and may not simply derive from suppression of an antiferromagnetic phase.

**ACKNOWLEDGEMENTS**

We thank B.C. Sales for fruitful the discussions and M. A. McGuire for assistance. The single crystal X-ray diffraction experiments were carried out with the assistance of R. Custelcean. This work was supported by the Department of Energy, Basic Energy Sciences, Materials Sciences and Engineering Division and Scientific User Facilities Division.

Corresponding author: saparovbi@ornl.gov